\documentclass[pra,twocolumn,showpacs,amsmath,amssymb]{revtex4}
\newcommand{\beq}[0]{\begin{equation}}
\newcommand{\eeq}[0]{\end{equation}}
\newcommand{\bw}[0]{\begin{widetext}}
\newcommand{\ew}[0]{\end{widetext}}
\newcommand{\bwn}[0]{\begin{widetext}\begin{eqnarray}}
\newcommand{\ewn}[0]{\end{eqnarray}\end{widetext}}
\newcommand{\beqn}[0]{\begin{eqnarray}}
\newcommand{\eeqn}[0]{\end{eqnarray}}

\newcommand{\np}[0]{{\it e.g. }}
\newcommand{\tzn}[0]{{\it i.e. }}

\newcommand{\proj}[1]{|#1\rangle \langle #1|}
\newcommand{\ket}[1]{|#1\rangle}
\newcommand{\bra}[1]{\langle #1 |}

\newcommand{\kan}[0]{\Lambda}
\newcommand{\tr}[0]{\mathrm{tr}\:}

\newcommand{\jedynka}[0]{\mathbb{I}}

\def\etal{{\it et al.}}

\def\calI{{\cal I}}

\usepackage{graphicx}
\usepackage{dcolumn}
\usepackage{bm}
\usepackage{bbm}
\begin{document}
\title{Reexamination of determinant based separability test for two qubits}
\author{Maciej Demianowicz}\email{maciej@mif.pg.gda.pl}
\affiliation{Faculty of Applied Physics and Mathematics,
Gda\'nsk University of Technology, PL-80-952 Gda\'nsk, Poland}
\affiliation{National Quantum Information Center of Gdansk, PL-81-824 Sopot, Poland}
\begin{abstract}
It was shown in [Augusiak \etal,\;Phys. Rev. A \textbf{77}, 030301(R) (2008)] that discrimination between entanglement and separability in a two qubit state can be achieved by a measurement of a single observable on four copies of it. Moreover, a pseudo entanglement monotone $\pi$ was proposed to quantify entanglement in such states. The main goal of the present paper is to show that close relationship between $\pi$ and concurrence reported there is a result of sharing the same underlying construction of a spin flipped matrix. We also show that monogamy of entanglement can be rephrased in terms of $\pi$ and prove the factorization law for $\pi$.
\end{abstract}
\pacs{03.67.Mn}
\maketitle

Entanglement, first recognized by Schrodinger, Einstein, Podolsky, and Rosen \cite{schrodingerepr}, lies at the heart of quantum information theory. With no doubt it is the most important resource of this rapidly developing branch of science and serves as the building brick for the huge number of information tasks, just to mention \np teleportation \cite{teleportation} and dense coding \cite{dense}. From this point of view full recognition of this 'spooky action at a distance' \cite{einstein} is fundamental to our understanding of quantum mechanics. Much effort has been put to
recognize its nature and not surprisingly the major progress has been achieved in the case of the simplest bipartite quantum states- states of two qubits. One of the most important qualitative results concerning such systems is the necessary and sufficient condition for inseparability- the celebrated Peres-Horodeckis criterion of non positive partial transposition \cite{ppt}. On the other side research towards quantitative description of entanglement of two qubit states has culminated in the introduction of entanglement measures, among which the most notable are entanglement of formation \cite{formation} and concurrence for whom
closed expressions has been found \cite{hillwoot}.  Unfortunately, both of them has still not been shown to be directly measurable and it is reasonable to conjecture that they are not in general. However, very recently it has been demonstrated that {\it single} collective measurement of the specially prepared observable on four copies of an unknown two qubit state can {\it unambiguously} discriminate between entanglement and separability, additionally {\it quantifying} to some extent entanglement contained in the system by providing sharp lower and upper bounds on concurrence \cite{nasza}.

In the present paper we continue research on the pseudo entanglement monotone $\pi$, which was introduced in Ref. \cite{nasza} for entanglement quantification purposes.

Let us start with the an introduction of necessary concepts.
Consider a two qubits mixed state $\rho_{AB}$. Define (conjugation in a standard basis) a spin flipped state \cite{hillwoot}
\beq
\tilde{\rho}_{AB}=\sigma_y \otimes \sigma_y \rho_{AB}^* \sigma_y \otimes \sigma_y.
\eeq
Let $\lambda_1 \ge \lambda_2 \ge \lambda_3\ge \lambda_4$ be the square roots of the eigenvalues of $\rho_{AB}\tilde{\rho}_{AB}:=M_{AB}$. Note that we can safely write inequalities since they are real (moreover they are nonnegative).
We define concurrence to be
\beq\label{concurrence}
C(\rho_{AB})=\max \{0,\lambda_1-\lambda_2 -\lambda_3-\lambda_4\}=:C_{AB}.
\eeq
Eligibility of such constructed quantity for being a good measure of entanglement is justified by the invariance of the eigenvalues $\lambda_i$ under local unitary operations and by the fact that $0\le C_{AB}\le 1$ with extreme values taken on separable and maximally entangled states respectively.
When $\rho_{AB}$ is a partial trace over $C$ from the tripartite pure qubit state $\proj{\psi_{ABC}}=:\psi_{ABC}$ there are only two
nonzero eigenvalues thus we have just $C_{AB}=\lambda_1-\lambda_2$.
We then also define  tangle \cite{ckw} to be
\beq\label{tangle}
\tau_{ABC}=4\lambda_1 \lambda_2.
\eeq
It was shown that eigenvalues of either of the matrices $M_{AB}$, $M_{BC}$, $M_{AC}$ can be used in the above.
These quantities can be  combined to give the so called monogamy relation \cite{ckw,osborn}
\beq\label{mono-ckw}
C_{AB}^2+ C_{BC}^2+\tau_{ABC}=C_{B(AC)}^2=4\det \rho_B,  \rho_B=\tr_A \;\rho_{AB}.
\eeq
Concurrence $C_{B(AC)}$ is the meaningful quantity since we consider pure state of three qubits thus effectively
$B(AC)$ is a two-qubit-like state. This relations provides an interpretation for tangle as a measure of tripartite correlations.

One also defines \cite{assist} concurrence of assistance $C^a$, which is the maximum over ensembles of average concurrence of pure states in the ensemble. In case of two qubits we have just $C^a=\lambda_1+\lambda_2+\lambda_3+\lambda_4$.

In Ref. \cite{nasza} it was shown that the separability of an unknown two qubit state $\rho$ can be unambiguously settled in a single collective measurement on four copies of this state, \tzn one needs at one time $\rho ^{\otimes 4}$. This was obtained on the basis of two facts: (i) partially transposed density matrix $\rho^{\Gamma}$ of an entangled two qubit state $\rho$ is full rank (has four non zero eigenvalues), (ii) there can be only one negative eigenvalue of $\rho^{\Gamma}$. The above led to the conclusion that it is sufficient to measure $\det \rho^{\Gamma}$ and the strict negativity of the latter indicates entanglement. The authors of the mentioned paper showed that indeed such a measurement is possible using a noiseless circuit \cite{nasza2}. They also proposed a simple alternative scheme to measure this determinant. The question of the usage of $\det \rho^{\Gamma}$ for quantitative description was further addressed. It was shown that the quantity, which we will call in this paper the {\it determinant-based measure},
\beqn\label{miara}
\pi(\rho)=\left\{\begin{array}{ccc}
               0 & \mathrm{for} & \det{\rho^{\Gamma}} \ge 0 \\
               2\sqrt[4]{|\det{\rho^{\Gamma}}|}  & \mathrm{for} & \det{\rho^{\Gamma}} < 0
                \end{array}.\right.
\eeqn
is a monotone under pure local operations preserving dimensions and classical communication and provides tight upper
and lower bounds on concurrence as follows
\beq\label{boundy}
C(\rho) \le \pi(\rho)\le \sqrt[4]{C(\rho) \left( \frac{C(\rho)+2}{3}\right)^3}.
\eeq
Normalization in Eq. (\ref{miara}) is chosen to impose agreement of determinant-based measure and concurrence on pure states.
From the above inequalities we also have immediate bounds for entanglement of formation $E_f$ \cite{hillwoot} as follows $E(r^{-1}(\pi(\rho)))\le E_f(\rho)\le E(\pi(\rho))$, where $E(x)=H(\frac{1+\sqrt{1-x^2}}{2})$ with $H(y)$ being the Shannon entropy of a probability distribution $(y,1-y)$ and $r(x)=\sqrt[4]{x \left( \frac{x+2}{3}\right)^3}$.
One can also prove that $\pi$ shares the nice property of being continuous in the input density operator \cite{contin}.

For the purpose of the present paper we propose the  extension of our definition for entanglement between
qubit $A$ and qubits $BC$ in a pure state $\psi_{ABC}$ to  $\pi_{A(BC)}\equiv 2\sqrt{\det \rho_A}$, \tzn we define it to be equal to $C_{A(BC)}$ on such states. Such extension is the most natural since we keep two most important properties of $\pi$: mentioned equality with concurrence and possibility of direct measurement.

\begin{figure}\label{fidelity-random}
\includegraphics[width=60mm,height=50mm]{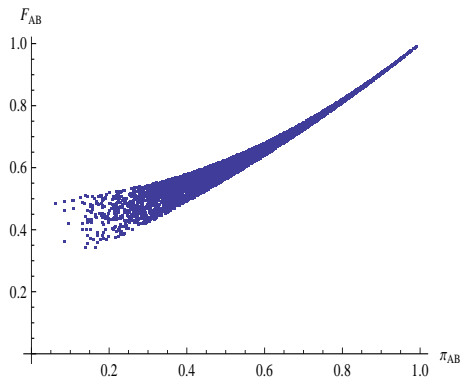}
\caption{Singlet fraction $F_{AB}$ {\it vs.} determinant-based measure $\pi_{AB}$ after action of the local channel}
\end{figure}

Let us now turn to the main body of the paper. We start with
considerations analogous to the one from Ref. \cite{ishizaka}, where the local unitary interaction of one part of maximally entangled state $\psi^+_{AB}$ with two level environment $E$ was considered, \tzn the global state after the evolution was $\ket{\psi_{ABE}}=\jedynka_A\otimes U_{BE}(\ket{\psi^{+}}_{AB}\otimes \ket{0}_E)$.

Its bipartite reductions of interest will be denoted $\rho_{AB}$ and $\rho_{AE}$.
In the mentioned paper the author showed by random sampling that there is no correlation between singlet fraction \cite{fraction} $F(\rho_{AB}):=F_{AB}$  after the action of the channel and concurrence $C_{AB}$ of the decohered state and showed analytically for the chosen class of channels that $F_{AB}=\frac{1}{4}(1+C_{AB})(1+\sqrt{1-C_{AE}^2})$. It turned out that this relation holds true for all channels, which was shown by random generation of channels.

We pursue the same approach using determinant-based measure instead of concurrence. First let us consider relation of $F_{AB}$ and $\pi_{AB}$ after the action of the random channel. The result is shown in the Fig.1., which was obtained by random generation of $10000$ $U_{BE}$'s \cite{zyczkowski}.

One can see that in our case there is some connection between these two quantities, however, still there is no analytical formula linking them. We will comment on this connection later when monogamy equality will be obtained.

Following Ref. \cite{ishizaka} let us consider a class of local channels implemented by unitaries defined by:
\beqn
\ket{00}_{BE}\to \sqrt{1-q} \ket{00}_{BE} + \sqrt{q} \ket{11}_{BE}\\\
\ket{10}_{BE}\to \sqrt{1-p} \ket{10}_{BE} + \sqrt{p} \ket{01}_{BE}.
\eeqn
For such channels we obtain the following (with previously established notation):
\beqn
\displaystyle
\pi_{AB}=\sqrt{|p+q-1|}\\\
\pi_{AE}=\sqrt{|p-q|}\\\
F_{AB}=\left\{\begin{array}{ccc}
               \frac{2-p-q+2\sqrt{(1-p)(1-q)}}{4} & \mathrm{for} & p+q-1 < 0 \\
               \frac{p+q+2\sqrt{pq}}{4}  & \mathrm{for} & p+q-1\ge 0
                \end{array}.\right.
\eeqn
Direct calculation reveals that \beq \displaystyle F_{AB}=\frac{1}{4}(1+\pi_{AB}^2)\left(1+\sqrt{1-\frac{\pi_{AE}^4}{(\pi_{AB}^2+1)^2}}\;\right).\eeq
As it was in the case of concurrence, the relation we obtained can be shown to hold for all channels and is independent of which maximally entangled state we choose to be the input. Note the close
resemblance of both forms.

Closed formula for singlet fraction using determinant-based measure $\pi$
 opens hope for a monogamy relation of entanglement in terms of it. In what follows we prove the existence of such equation.

Consider a pure state of three qubits $\psi_{ABC}$. As it was shown \cite{3szmit}, as far as the entanglement properties are considered, such state can be parameterized by five real numbers as
\beqn
\ket{\psi_{ABC}}=\gamma_0 \ket{000}+\gamma_1\mathrm{e}^{\mathrm{i}\varphi}\ket{100}+\gamma_2\ket{101}+\nonumber \\ \gamma_3\ket{110}+\gamma_4\ket{111}
\eeqn
with $\gamma_i\ge 0$, $\sum_i \gamma_i^2=1$, and $\varphi\in \langle 0,\pi \rangle$.
From this we obtain eigenvalues of the matrix $M_{AB}$ and determinant of the partially transposed matrix of the reduced state of qubits $A$ and $B$
\beqn
&&\lambda_1^2 =\gamma_0^2 \left(2\gamma_3^2+\gamma_4+2 \gamma_3\sqrt{\gamma_3^2+\gamma_4^2}\right),  \\\
&&\lambda_2^2 =\gamma_0^2 \left(2\gamma_3^2+\gamma_4-2 \gamma_3\sqrt{\gamma_3^2+\gamma_4^2}\right),\\\
&&\det \rho_{AB}^{\Gamma}=-\gamma_0^4\gamma_3^2(\gamma_3^2+\gamma_4^2),
\eeqn
which immediately yields
\beq
\pi_{AB}=\sqrt{\lambda_1^2-\lambda_2^2}.
\eeq
Recalling Eqs. (\ref{concurrence}) and (\ref{tangle}) we obtain analytical relationship between $\pi_{AB}$, $C_{AB}$, and $\tau_{ABC}$ in a pure three qubit state
\beq\label{pi}
\pi_{AB}=\sqrt{C_{AB}\sqrt{C_{AB}^2+\tau_{ABC}}}.
\eeq
This can be put into a nice compact form
\beq
\pi_{AB}=\sqrt{C_{AB}C^a_{AB}}\;,
\eeq
which means that in case of rank two states the determinant-based measure is the geometric mean value of concurrence and concurrence of assistance. We also conclude that the bound in Eq. (\ref{boundy}) can be tightened for such states to obtain
\beq
\pi_{AB}\le \sqrt{C_{AB}}.
\eeq

Eq. (\ref{pi}) leads us to a simple corollary stating that for a given pure three qubit state $\psi_{ABC}$ one has $\pi_{AB}=C_{AB}$ if and only if $C_{AB}=0$ or $\tau_{ABC}=0$. With the results of \cite{corollary} this means that both measures agree when $\psi_{ABC}$ is either of the following classes: $GHZ$ with separable reduction $AB$, $W$, biseparable, or product.

One can now argue that the pattern in the plot of singlet fraction (Fig.1.) is the result of quantifying to some extent tripartite correlations by the determinant-based measure.

Now let us reverse (\ref{pi}) to get
\beq
C_{AB}^2=\frac{-\tau_{ABC}+\sqrt{\tau_{ABC}^2+4\pi_{AB}^4}}{2}.
\eeq
Inserting this into Eq. (\ref{mono-ckw}) one obtains the advertised elegant monogamy relation in terms of the determinant-based measure
\beq\label{monogamia}\displaystyle
\sqrt{(\frac{\tau_{ABC}}{2})^2+\pi_{AB}^4}+\sqrt{(\frac{\tau_{ABC}}{2})^2+\pi_{BC}^4}=\pi^2_{B(AC)}.
\eeq
This gives also the recipe to measure tangle on ten copies of the state relying directly on the measurements of determinants of two partially transposed density matrices (four plus four copies) and the determinant of the reduced qubit density matrix (two copies). The question of optimality of such measurement is beyond the scope of this paper (see \cite{carteret}).

Now we will argue that the determinant-based measure $\pi$ in a general case of mixed states of arbitrary rank is an analytical function of the eigenvalues of the matrix $M$.
Consider states
\beqn
\rho_{Bdiag}=p_1 \psi_+ +p_2 \psi_- +p_3 \phi_+ + p_4 \phi_-,
\eeqn
which are diagonal in the Bell basis $\ket{\psi_{\pm}}=\frac{1}{\sqrt{2}}(\ket{01}\pm \ket{10})$, $\ket{\phi_{\pm}}=\frac{1}{\sqrt{2}}(\ket{00}\pm \ket{11})$. Such states are entangled iff one of the probabilities is larger than $\frac{1}{2}$. W.l.o.g. assume that it holds for $p_1$.
One has then
\bwn
\pi(\rho_{Bdiag})=\sqrt[4]{|-p_1+p_2+p_3+p_4|(p_1-p_2+p_3+p_4)(p_1+p_2-p_3+p_4)(p_1+p_2+p_3-p_4)}.
\ewn
We also easily compute that $\lambda_i=p_i$. Motivated by the form of $\pi$ for $\rho_{Bdiag}$ we further define
\beqn
C_1=\max \{0,\lambda_1-\lambda_2 -\lambda_3-\lambda_4\}\equiv C,\\\
C_2=\lambda_1-\lambda_2 +\lambda_3+\lambda_4,\\\
C_3=\lambda_1+\lambda_2 -\lambda_3+\lambda_4,\\\
C_4=\lambda_1+\lambda_2 +\lambda_3-\lambda_4
\eeqn
and
\beqn
\hat{\pi}(\rho)=\sqrt[4]{C_1 C_2 C_3 C_4}.
\eeqn
Notice that $C_2,C_3,C_4$ are always non negative.

The main result of this part of the paper is the following.
\newline\noindent
\textit{Theorem 1.} For any two qubit state $\rho$ one has
\beq\label{formula}
\pi(\rho)=\hat{\pi}(\rho).
\eeq
{\it Proof.} Let $A$ and $B$ be non singular local filters. The initial state $\varrho_1$ after the transformation under these filters is $\varrho_2=(1/p) A\otimes B \varrho_1 A^{\dagger}\otimes B^{\dagger}$, where $p=tr A\otimes B \varrho_1 A^{\dagger}\otimes B^{\dagger}$. It follows from the results of Ref. [20] that $C_i(\varrho_2)=(|\det AB|/p)C_i(\varrho_1)$. Moreover, it holds true that $\pi(\varrho_2)=(|\det AB|/p)\pi(\varrho_1)$ [8].  Assume now that $\varrho$ is a rank $4$ state.
It was shown [20] that such states can be reversibly obtained with $A$, $B$ from a Bell diagonal state $\varrho_{Bdiag}$. As we already know, the assertion of the theorem is true on the latter. We thus have $\pi(\varrho)=(|\det AB|/p)\sqrt[4]{\Pi_i C_i(\varrho_{Bdiag})}$ and because of the transformation rule for $C_i$ it follows that for full rank states it holds $\pi(\varrho)=\hat{\pi}(\varrho)$. For singular states, we can take their full rank perturbations $\sigma_{\epsilon}=\epsilon \varrho+(1-\epsilon)\mathbb{I}/2$. Then, $\pi(\sigma_{\epsilon})=\hat{\pi}(\sigma_{\epsilon})$ and we can apply the preceding argument and take the limit $\epsilon\to 0$. The result then follows from continuity of $\pi$ and $C_i$. $\blacksquare$

We see that $\pi$ can be regarded as some kind of symmetrization of concurrence allowing for experimental direct accessibility. Natural question is to what extent determinant-based measure quantifies also tripartite correlations in the general case. Unfortunately we have not been able to find definite answer so far \cite{step}.

At the end, we prove the factorization law, which was originally stated for concurrence \cite{konrad}.

\noindent
\textit{Theorem 2.} Determinant-based measure $\pi$ obeys the factorization law \tzn for an arbitrary channel $\kan$, pure state $\phi$, and a Bell state $\psi_+$ it holds
\beq
\pi(\calI\otimes \kan (\phi))=\pi (\calI\otimes \kan (\psi_+))\pi(\phi).
\eeq
\textit{Proof.} The assertion is trivially true for separable $\mathcal{I}\otimes \Lambda (\psi_+)$ ({\it i.e.}, when $\Lambda$ is entanglement breaking) so we may assume entanglement of the latter. Any state $\ket{\phi}$ can be written as $A\otimes \mathbb{I}(\ket{\psi_+})$ with $tr A^{\dagger}A=2$. We then have $\pi(\mathcal{I}\otimes \Lambda (\phi))=2\sqrt[4]{|\det[ \mathcal{I}\otimes \Lambda (A\otimes \mathbb{I}(\psi_+)A^{\dagger}\otimes \mathbb{I})]^{\Gamma_B}|}=2\sqrt[4]{|\det[A\otimes \mathbb{I}(\varrho_{\Lambda}^{\Gamma_B})A^{\dagger}\otimes \mathbb{I}]|}$, where $\varrho_{\Lambda}=\mathcal{I}\otimes \Lambda(\psi_+)$ and we have used the fact that $(X\otimes \mathbb{I} \varrho Y\otimes \mathbb{I})^{\Gamma_{B}}=X \otimes \mathbb{I} \varrho^{\Gamma_B} Y\otimes \mathbb{I}$. Using now the following property of the determinant $\det XY=\det X \det Y$ and the fact that $\pi(\phi)=|\det A|$ [8] we arrive at $\pi(\mathcal{I}\otimes \Lambda (\phi))=\pi(\phi)\pi(\mathcal{I}\otimes \Lambda (\psi_+))$ which is the desired. $\blacksquare$

Thus the determinant-based measure provides {\it factorizable} measurable bound on concurrence (see \cite{factor} for recent effort in this direction).

We have not been able to find an analytical proof of the extension of the factorization law to the mixed state domain, as it was in \cite{konrad}, nevertheless, by random sampling, we have verified that such an extension is indeed valid, that is $\pi(\mathcal{I}\otimes \Lambda (\varrho))\le\pi(\varrho)\pi(\mathcal{I}\otimes \Lambda (\phi_+))$

In conclusion, we have provided monogamy relation for entanglement quantified by determinant based measure $\pi$. As a byproduct we obtained explicit formulas for the latter in terms of other entanglement quantities. We showed that
close relation with concurrence is the result of bearing the similar construction in its roots.
 We also provided evidence that the disagreement of $\pi$ and $C$ on general mixed states stems from the fact that $\pi$ quantifies to some extent both bipartite and tripartite correlations. The  natural question motivated by the result of the present paper is about possibility of constructing other measurable quantifiers of entanglement, which are based on the analogous procedure and provide better bounds on concurrence of an unknown state. We hope our results will stimulate research on this topic and will provide some tools for improved understanding of two qubits entanglement.
 The issue of using the determinant-based measure for detecting and quantifying entanglement in higher dimensional systems is the subject of the ongoing research \cite{prep}.

Discussions with R. Augusiak and P. Horodecki are gratefully acknowledged. The author is supported by Ministerstwo Nauki i Szkolnictwa Wyzszego grant No. N N202 191734.


\begin{thebibliography}{0}
\bibitem{contin} Assume $||\rho-\sigma||_1\le \epsilon$, where $||\cdot||_1$ denotes the trace norm. Recall that for any hermitian operator on $d\otimes d$ it holds $||M^{\Gamma_A}||_1\le d||M||_1$. From this relation we immediately obtain that  $||\rho^{\Gamma_A}-\sigma^{\Gamma_A}||_1\le 2\epsilon$.
    Let $\lambda_i^\downarrow$ and $\sigma_j^\downarrow$ be the eigenvalues of $\rho^{\Gamma_A}$ and $\sigma^{\Gamma_A}$ ordered decreasingly. Then obviously $\forall_i |\lambda_i^\downarrow-\sigma_i^\downarrow|\le 2\epsilon$. From this we obtain $|\det \rho^{\Gamma_A}-\det \sigma^{\Gamma_A}|\le \eta$ with $\eta \to 0$ when $\epsilon \to 0$.
\bibitem{fraction} It is defined by $F(\rho)=\max_{\psi_{m}}\bra{\psi_m}\rho \ket{\psi_m}$ with $\psi_m$ maximally entangled state.
\end{thebibliography}
\end{document}